\documentclass[prb,twocolumn,showpacs,superscriptaddress]{revtex4}

\usepackage{graphicx}
\usepackage{dcolumn}
\usepackage{bm}
\usepackage{color} 

\begin{document}

\title{Berry phase and Rashba fields in quantum rings in tilted magnetic field}

\author{V. Lopes-Oliveira}
\affiliation{Departamento de F\'{\i}sica, Universidade Federal de
S\~{a}o Carlos, 13565-905, S\~{a}o Carlos, S\~{a}o Paulo, Brazil}
\affiliation{Department of Physics and Astronomy and Nanoscale and Quantum Phenomena Institute, Ohio University,
45701-2979, Athens, Ohio, USA }
\author{L. K. Castelano}
\affiliation{Departamento de F\'{\i}sica, Universidade Federal de
S\~{a}o Carlos, 13565-905, S\~{a}o Carlos, S\~{a}o Paulo, Brazil}
\author{G. E. Marques}
\affiliation{Departamento de F\'{\i}sica, Universidade Federal de
S\~{a}o Carlos, 13565-905, S\~{a}o Carlos, S\~{a}o Paulo, Brazil}
\author{S. E. Ulloa}
\affiliation{Department of Physics and Astronomy and Nanoscale and Quantum Phenomena Institute, Ohio University,
45701-2979, Athens, Ohio, USA }
\author{V. Lopez-Richard}
\affiliation{Departamento de F\'{\i}sica, Universidade Federal de S\~{a}o Carlos, 13565-905, S\~{a}o Carlos, S\~{a}o Paulo, Brazil}

\begin{abstract}

We study the role of different orientations of an applied magnetic
field as well as the interplay of structural asymmetries on the characteristics of eigenstates in a
quantum ring system. We use a nearly analytical model description of
the quantum ring, which allows for a thorough study of elliptical
deformations and their influence on the spin content and Berry phase of different
quantum states. The diamagnetic shift and
Zeeman interaction compete with the Rashba spin-orbit interaction, induced by confinement
asymmetries and external electric fields, to
change spin textures of the different states. Smooth variations
in the Berry phase are observed for symmetric quantum rings as function of applied magnetic fields.
Interestingly, we find that asymmetries induce nontrivial Berry
phases, suggesting that defects in realistic structures would
facilitate the observation of geometric phases.
\end{abstract}


\maketitle

\section{Introduction}

The phase acquired when a system is
subjected to a cyclic adiabatic process, as described by Berry and others, \cite{Berry,Wilczek,XiaoRMP}
contains information on the geometrical properties of the parameter space over which the system is defined.
In a spatially extended and multiply connected quantum system, this phase conveys nonlocal information
on the system and possible net fluxes akin to the Aharonov-Bohm phase. \cite{ABref}
As such, it is attractive to develop experimental probes to measure this Berry phase, as well as
theoretical models that connect its behavior to microscopic information or external fields.
The geometric Berry phase has indeed played a fundamental role in understanding the
behavior of a variety of systems and phenomena. \cite{Wilczek,XiaoRMP,Berry2}

In mesoscopic systems, the Berry phase in electronic states has been explored
by transport experiments in different systems,  providing a unique window into microscopic fields
and spin textures that arise from the interplay of external fields, as well as intrinsic spin-orbit
effects in structures defined on semiconductors. \cite{Berryorigin,Morpurgo,Shayegan,Nitta,ThomasRing}
More recently, transport  experiments have demonstrated that it is possible to
control the geometric phase of electrons by the application of in-plane fields
in semiconductor quantum rings built on InGaAs structures. \cite{1}

Motivated by these
experiments, we present here an analysis of the influence of
magnetic field orientation and intensity on the Berry phase
experienced by electrons in a realistic quantum nanoring structure.
As we will describe, the modulation of the geometric
phase can arise from the symmetry reduction in the
confinement potential or the competition between the external
magnetic field and the intrinsic field arising from spin-orbit coupling effects.
As such, this study addresses the link between spatial symmetry and spin
properties, and the possible tuning of the geometrical phase by varying the intensity
and/or orientation of an external magnetic field.

To this end, we use an effective mass description of the conduction
band, and incorporate the effects
of confinement asymmetry for electrons in a realistic nanoring, as well as the resulting Rashba
spin-orbit coupling (SOC) fields arising from confinement and external fields.
By studying spin maps for angle and magnetic field intensities,
we gain insights into the competition between
different energy scales and how they impact the Berry phase associated
with each electronic state.
As level mixing is enhanced under near resonant conditions, one anticipates
interesting behavior at the anticrossing regions
produced for example by varying magnetic field dependence in a given structure.
There are pronounced
spatial asymmetry effects in the angular momentum and spin character of different states, as one
would expect.  These
asymmetries, introduced or enhanced by shape anisotropies and confinement
potential in the rings, are found to play an important role in determining the Berry phase of
the different states. We also find that effects of varying magnetic
field tilt angle and intensity, as well as SOC, are reflected in the Berry phase
and associated spin texture.
The substantial phase modulation observed in the
lower energy level manifold can be monitored and exploited in
transport experiments.

The remainder of the paper is organized as follows: Sec.\ \ref{sec:model} describes the
theoretical model used, as well as the different quantities used to characterize the states, including
the Berry phase and spin density maps. Section \ref{sec:results} is devoted to the discussion of
the main results, while concluding remarks are presented in Sec.\ \ref{sec:disc}.

\section{Model} \label{sec:model}

\begin{figure}
\includegraphics[scale=1.0]{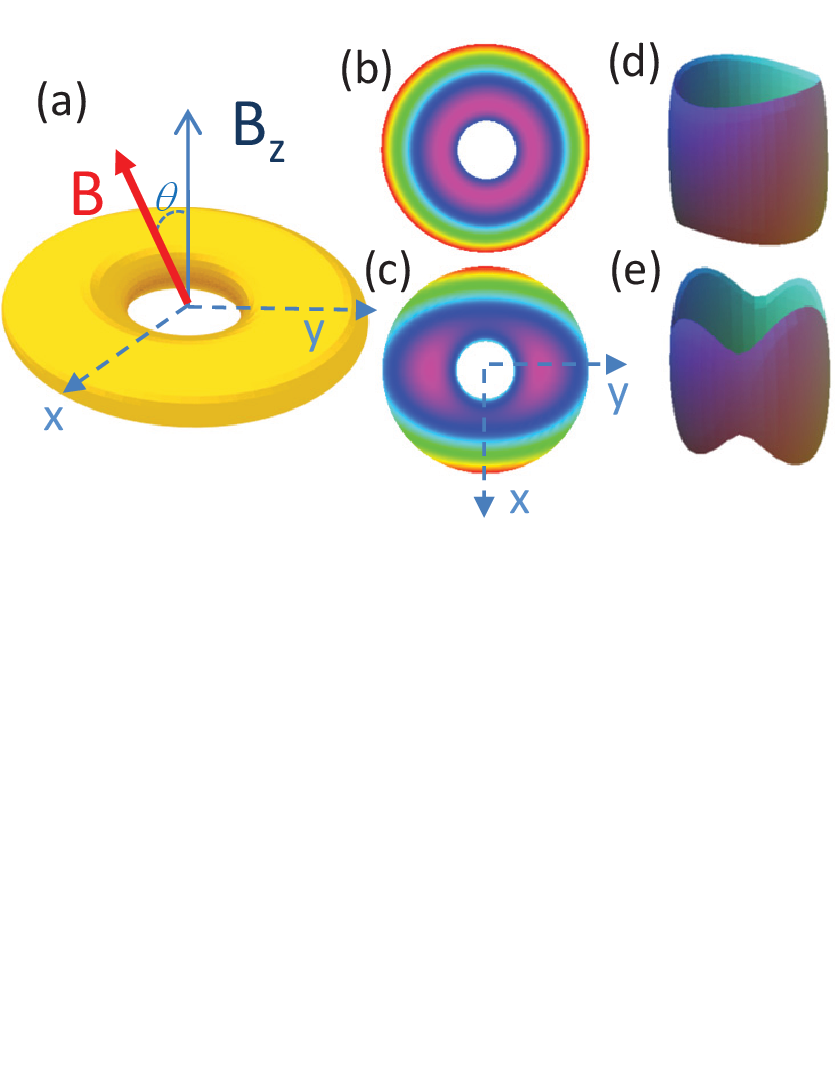}
\caption{(a) Magnetic field orientation and coordinate system. Potential profile
maps for (b) the circularly symmetric and (c) the elliptically deformed ring; purple regions
show lowest energy values of the confinement potential. Electronic orbital for an
excited state of an elliptical ring ($\delta=2$ meV), in the presence of a Rashba
field ($F=100$ kV/cm) in a magnetic field  ($B=2.375$ T) at
different tilt angles, (d) $\theta=0^{\circ}$ and (e) $\theta=60^{\circ}$.}
\label{fig1}
\end{figure}

The system under investigation consists of a quantum ring in the
presence of an external tilted magnetic field, as shown in Fig.\
\ref{fig1}(a). The confinement potential, $V(\vec{r}) = V_\rho +
V_z$, with a general elliptical ring shape is characterized by
planar and vertical contributions, \cite{2}
\begin{eqnarray}\label{profile}
V_\rho&=&\frac{a_1}{\rho^2}+a_{2}\rho^2-2\sqrt{a_{1}a_2}+\delta\rho^2\cos^2{\varphi} \,, \\
\label{profilez}
V_z&=&\left\{
\begin{array}{ll}
eFz, \,\,\, &0<z\leq L_{z} \\
\infty, &{\rm otherwise} \, ,
\end{array}%
\right.
\end{eqnarray}
where $L_{z}$ is the height/thickness of the quantum well in which
the ring is defined by the lateral potential $V_\rho$.  $F$ is an
external electric field applied along the ring axis, which gives
rise to a Rashba SOC, in addition to shifting and spatially
deforming the eigenstates. The radial potential without the
$\delta$-term defines a ring with circular symmetry and minimum
($V_{\rho = R}=0$) at $R=(a_1/a_2)^{1/4}$. The $\delta$-term
describes an elliptical deformation of the confinement potential,
with eccentricity given by $e=\sqrt{1-a_{2}/(a_{2}+\delta)}$ in the limit of large $\rho$.
Panels (b) and (c) in Fig.\ \ref{fig1} represent
potential profile maps used to simulate circularly symmetric
($\delta=0$) and eccentric ($\delta\neq0$) rings, respectively. This
potential profile allows for analytical solution of the quantum ring
spectrum and corresponding wave functions in the $x$-$y$ plane in
the circularly symmetric case, in terms of hypergeometric functions
and angular momentum components.\cite{2} The electronic structure
calculations are performed using the effective mass approximation
and consider a single quantum level along the $z$-axis (strong
quantum well confinement). This model has been used successfully to
describe quantum rings in magneto photoluminescence experiments.
\cite{2,4}  It is extended here to include SOC effects and tilted magnetic fields.

\subsection{Tilted Magnetic Field}

In the presence of a magnetic field $\vec{B}=\hat{x} B_x + \hat{z}
B_z = \hat{x} B\sin \theta+ \hat{z}B\cos \theta$, the vector
potential can be written as
\begin{equation}\label{potvec}
\vec{A}=\frac{B_z}{2}\rho\hat{\varphi}-B_x{z} \left ( \hat{\rho} \sin \varphi +\hat{\varphi} \cos \varphi \right) \, .
\end{equation}
It is important to note that small variations of the magnetic field
angle $\theta$ induce considerable changes in the electronic
structure, as the tilted magnetic field and ring asymmetry ($\delta
\neq 0$) couples angular and radial degrees of freedom. In the
absence of SOC, the system is described by the Hamiltonian,
\begin{equation}\label{Hamiltonian}
\emph{H}=\frac{1}{2\mu^*} \left(\vec{p}-e\vec{A} \right)^2+V(\vec{r})+\frac{1}{2}g\mu_B\vec{B}\cdot\vec{\sigma} \, ,
\end{equation}
where the third term is the Zeeman
interaction and $\vec{\sigma}=(\sigma_{x},\sigma_{y},\sigma_{z})$
are the Pauli matrices. Eq.\ (\ref{Hamiltonian}) can be separated into
three parts, $\emph{H}=H_{B_z}+H_{B_x}+H_{Z_x}$. The contribution due
to the perpendicular component of the magnetic field ($B_z$) is
given by
\begin{eqnarray}
H_{B_z}&=&-\frac{\hbar^2}{2\mu^*}\left[\frac{1}{\rho}\frac{\partial}{\partial\rho}\left(\rho\frac{\partial}{\partial\rho}\right)+\frac{1}{\rho^2}\frac{\partial^2}{\partial\varphi^2}+\frac{\partial^2}{\partial{z^2}}\right]\nonumber\\
&+&\frac{ie\hbar{B_z}}{2\mu^*}\frac{\partial}{\partial\varphi}+\frac{e^2B_z^2\rho^2}{8\mu^*}+V(\vec{r})\\
&+& \frac{g\mu_B}{2}B_z\sigma_z \, .\nonumber
\end{eqnarray}
The eigenfunctions of the circularly symmetric problem ($\delta=0$)
in the presence of the $B_z$ component,
$\Phi_{lmn}(z,\rho,\varphi)$, are used as the basis set to expand
the eigenstates for a general tilted field direction, under SOC, and
a general elliptical deformation. A general wavefunction can be
written as
\begin{eqnarray}
\Psi&=&\sum_{l,m,n} \left( C_{lmn}^{\uparrow} \mid\uparrow\rangle%
+ C_{lmn}^{\downarrow} \mid\downarrow\rangle \right)\Phi_{lmn} \, ,
\end{eqnarray}
where the spatial dependence has been omitted for simplicity. The
term due to the in-plane component of the magnetic field is
\begin{eqnarray}
H_{B_x}&=&-\frac{ie\hbar{z}B_x}{\mu^*}\left(\sin \varphi \frac{\partial}{\partial\rho}+\frac{\cos \varphi }{\rho}\frac{\partial}{\partial\varphi}\right)\nonumber\\
&+&\frac{e^2}{2\mu^*}\left(B_x^2z^2-B_zB_xz\rho\cos \varphi \right)
\, .
\end{eqnarray}
and the respective Zeeman contribution can be written
as\cite{zeemansplitting}
\begin{equation}
H_{Z_x}=\frac{1}{4}g\mu_B B_x(\sigma^{+}+\sigma^{-}) \, ,
\end{equation}
where $\sigma^{\pm}=\sigma_x \pm i\sigma_y$.

\subsection{Spin-Orbit Coupling}

The presence of spin-orbit coupling in the host semiconductor is
also considered for the tilted magnetic field case.  The SOC in the
presence of system inversion asymmetry can be written in terms of
the field associated with the confinement potential, $\nabla
V(\textbf{r})$, as \cite{3}
\begin{equation}\label{cartezianas}
H_{SIA}=\frac{\alpha_{s}}{\hbar} \vec{\sigma}\cdot \left( \nabla V \times (\vec{p} -e \vec{A}) \right) \, ,
\end{equation}
where $\alpha_s$ characterizes the strength of the
SOC in the host semiconductor.  This can be decomposed in cylindrical
coordinates into four terms, $H_{SIA}$=$H_{SIA}^{D} + H_R + H_K + H^{TF}_{SIA}$, where \cite{3}
\begin{eqnarray}\label{SIA}
H_{SIA}^D&=&\alpha_{s}\sigma_z\left\{\frac{\partial{V}}{\partial\rho}\left[-\frac{i}{\rho}\frac{\partial}{\partial\varphi}+%
\frac{eB_{z}}{2\hbar}\rho\right]+\frac{i}{\rho}\frac{\partial{V}}{\partial\varphi}\frac{\partial}{\partial\rho}\right.\nonumber\\
&+&\left.\frac{i}{\rho^2}\frac{\partial{V}}{\partial\varphi}\right\}
\, ,
\end{eqnarray}
\begin{eqnarray}\label{HR}
H_{R}&=&-\alpha_{s}\frac{\partial{V}}{\partial{z}}\left\{\sigma^{+}\left[e^{-i\varphi}\left(\frac{\partial}{\partial\rho}-%
\frac{i}{\rho}\frac{\partial}{\partial\varphi}+\frac{eB_{z}}{2\hbar}\rho+\frac{1}{\rho}\right)\right]\right.\nonumber\\
&-&\left.\sigma^{-}\left[e^{i\varphi}\left(\frac{\partial}{\partial\rho}+\frac{i}{\rho}\frac{\partial}{\partial\varphi}-%
\frac{eB_{z}}{2\hbar}\rho+\frac{1}{\rho}\right)\right]\right\} \, ,
\end{eqnarray}
and $H_{K}=0$ because $\langle{k_z}\rangle \simeq 0$. $H_{SIA}^{D}$
is the spin-diagonal contribution due to the confinement, while the Rashba
term $H_R$ is associated with the perpendicular electric field in
the well, $\partial V/ \partial z=eF$.  For a tilted magnetic field,
the last term in $H_{SIA}$ is given by
\begin{eqnarray}
H_{SIA}^{TF}&=&\alpha_s\frac{ezB_x}{\hbar}\left(\frac{\partial{V}}{\partial{z}}\sigma_x-\frac{\partial{V}}{\partial{x}}\sigma_z\right).
\end{eqnarray}

The SOC mixes states depending on their spin
component, following effective ``selection rules'' that select
specific angular momentum quantum numbers according to the corresponding interaction term. \cite{3}
As a consequence, the spin and angular momentum content of each
state become hybrids or mixtures that change with field orientation and magnitude. For large
magnetic fields, the Zeeman energy dominates and eventually
polarizes spins along $\vec{B}$.  Mixing is of course more
noticeable near spectrum degeneracies, as we will see below.

\subsection{Spin Content and Berry Phase}

We characterize the spin content of different eigenstates by
analyzing the expectation value for the different components. In
particular, we define the spin projection respect to the $z$-axis,
$\theta_s$, in terms of projections along and perpendicular to the
plane,
\begin{eqnarray}
\langle \sigma^{+}\rangle&=&\sum_{j}{C_{j}^{\uparrow*}C_{j}^{\downarrow}} \, , \nonumber \\
\langle\sigma_{z}\rangle&=&\sum_{j}\left(|C_{j}^\uparrow|^2-|C_{j}^\downarrow|^2
\right) \,
\end{eqnarray}
(where $j =\{n,l,m\}$ in all sums), so that
\begin{equation}
\theta_s=\arctan \frac{\langle\sigma_{z}\rangle}{\langle\sigma^{+}\rangle} +
\frac{\pi}{2}\left(1- {\rm sgn} \langle\sigma^{+}\rangle \right) \, .
\end{equation}

We also explore the spatial variation in the spin orientation (`spin
texture') for each state, which is related to the vector spin
density, whose components are given by
\begin{eqnarray}
S_x(\vec{r})&=& \sum_{j,j'} \Phi_{j'}^* (\vec{r}) \left( C_{j'}^{\uparrow *} C_j^\downarrow + %
C_{j'}^{\downarrow *} C_j^\uparrow \right) \Phi_j (\vec{r}) \nonumber \\
S_y(\vec{r})&=& -i\sum_{j,j'} \Phi_{j'}^* (\vec{r}) \left( C_{j'}^{\uparrow *} C_j^\downarrow - 
C_{j'}^{\downarrow *} C_j^\uparrow \right) \Phi_j (\vec{r}) \nonumber \\
S_z(\vec{r})&=&\sum_{j,j'} \Phi_{j'}^* (\vec{r}) \left( C_{j'}^{\uparrow *} C_j^\uparrow  -%
C_{j'}^{\downarrow *} C_j^\downarrow \right) \Phi_j (\vec{r}) \, .
\end{eqnarray}

The Berry phase is an interesting quantity that characterizes the different eigenstates,
especially as it incorporates the effects of external and SOC fields, and the
influence of geometrical confinement.
The Berry phase of a given eigenstate $\alpha$ is defined as \cite{XiaoRMP}
\begin{equation}\label{BPequation}
\Theta_{\alpha}=i \int_0^{2\pi} \langle{\Psi_{\alpha}|\frac{\partial}{\partial{\hat\varphi}}|%
\Psi_{\alpha}\rangle} \, d \hat\varphi \, ,
\end{equation}
where $\hat\varphi$ parametrizes a cyclic adiabatic process; we consider
here a closed path around the ring, so that $\Psi_\alpha (\varphi) \rightarrow
\Psi_\alpha(\varphi + \hat\varphi)$.  Different experiments would
probe the Berry phases in different fashion, depending on the
measurement design.  Transport phase measurements, for example,
would result in a mostly additive contribution of various channels
involved in the conductance signal, i.e., those close to the Fermi
energy.  We illustrate the effect of cumulative phase by considering
the total Berry phase for a collection of states, defined over a
certain `occupation' in the ring (defined once such structure is
connected to current reservoirs and a bias window is defined).

\section{Results} \label{sec:results}

The calculation of the spectrum in the ring utilizes the full diagonalization of the Hamiltonian written in the
basis that considers a sufficiently large Hilbert space, truncated to the desired accuracy.  We typically consider 11 
eigenstates, with angular momentum $|m| < 5$ 
for each spin orientation.  These are found sufficient for
convergence in the entire field and parameter range considered in
this work. \cite{2}

\begin{figure}[tbp]
\includegraphics[scale=1.0]{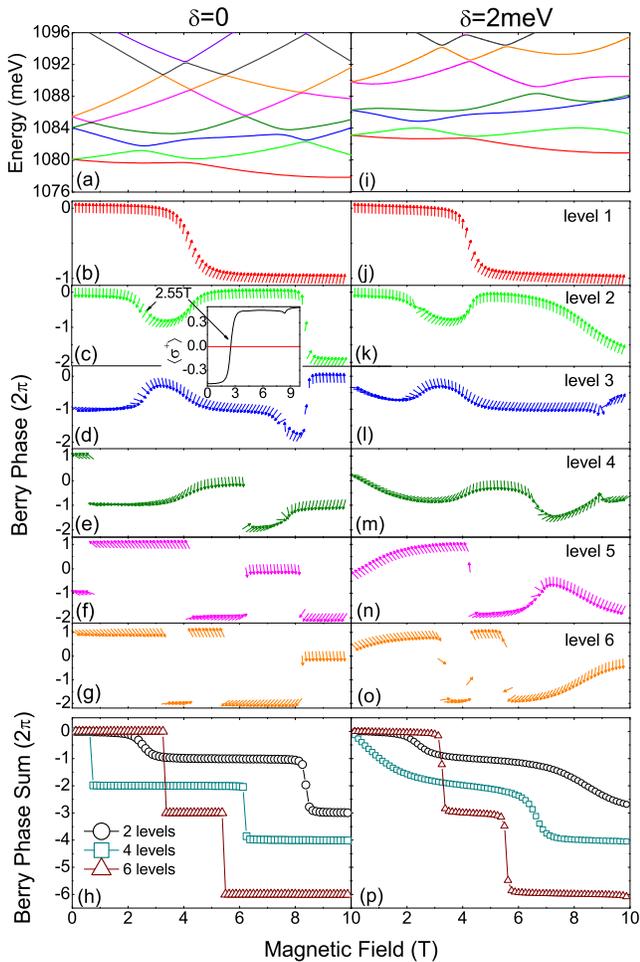}
\caption{Electronic structure for quantum rings in
a magnetic field at fixed tilt angle, $\theta=60^{\circ}$, and Rashba field $F=100$ kV/cm,
as a function of the total magnetic field strength for:
(a) symmetric ($\delta=0$) and (i) asymmetric ($\delta=2$ meV) ring.
The Berry phase for different levels for $\delta=0$ is shown in panels on the left column,
(b) through (g); and for $\delta=2$ meV on the right column, (j) through (o).  The
cumulative Berry phase for different occupation numbers is shown in panel (h) for the
symmetric ring, and panel (p) for the asymmetric ring. } \label{fig2}
\end{figure}

In what follows, we will assume parameters corresponding to InAs,
with electron effective mass $\mu^{*}=0.0229m_0$, Land\'{e} g-factor
$g=-14.9$, and SOC parameter, $\alpha_s=117.1$ \AA$^2$, taken from
the literature. \cite{Winkler}

\begin{figure}[tbp]
\includegraphics[scale=1.0]{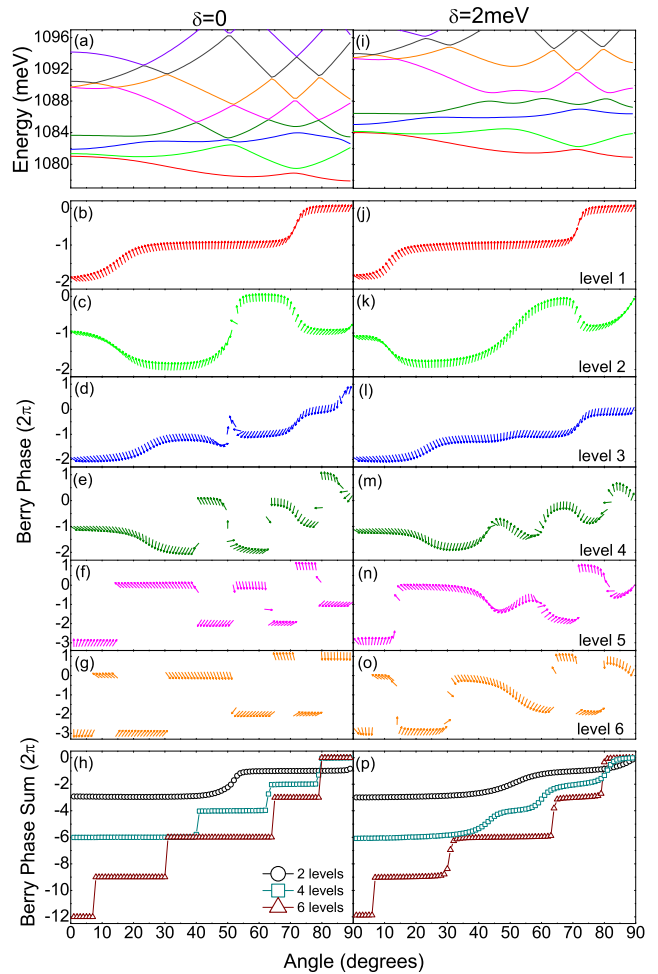}
\caption{Electronic structure for quantum rings under
fixed magnetic and Rashba fields, $B=6.625$ T and $F=100$ kV/cm, as a function of the magnetic field
tilt angle $\theta$
for: (a) symmetric ($\delta=0$) and (i) asymmetric ($\delta=2$ meV) ring.  Berry phases
for different states in both rings are shown in the panels below.  The cumulative Berry
phase for different occupations is shown in panels (h) and (p), for the symmetric and
asymmetric rings, respectively.  } \label{fig3}
\end{figure}

Figures \ref{fig2} and \ref{fig3} show the electronic structure and
the Berry phases for the lower energy manifold in both symmetric and
asymmetric rings. In Fig.\ \ref{fig2} we plot the energy levels and
corresponding phases as function of the total magnetic field
amplitude at a fixed angle, $\theta=60^{\circ}$, while Fig.\
\ref{fig3} shows results for a fixed intensity of the magnetic
field, $B=6.625$ T, as a function of the orientation $\theta$.
The Berry phases of the lowest six levels are displayed in the
figures \ref{fig2} and \ref{fig3}, along with the corresponding mean
spin orientations. The arrows along the different Berry phase curves
indicate the spin orientation, with $\theta_s$ as defined above:  An
upwards/downwards arrow in these curves, $\theta_s=\pm \pi/2$,
indicates a spin aligned along the $\pm z$-axis, while a horizontal
arrow indicates a spin lying on the $x$-$y$ plane.

The results on the left panels of both figures are for a circularly
symmetric quantum ring.  For small magnetic field, the two lowest
energy states exhibit spin alignment along $\pm z$-axis, as shown in
Fig.\ \ref{fig2}(b) and (c). On the other hand, the next four
levels (Fig.\ \ref{fig2}(d)-(g)) are aligned mostly on the
plane due to the spin mixing caused by SOC.\@ Notice that at high
values of magnetic field levels become essentially aligned with
(Fig.\ \ref{fig2}(b)-(d)) or against (Fig.\
\ref{fig2}(e)-(g)) the magnetic field, as the Zeeman
energy dominates over the SOC.\@ The evolution of spin orientation for
each level is strongly influenced by the anticrossings with other
levels, as one would expect. Moreover, anticrossings also affect the
Berry phase of states, causing a smooth variation with large
amplitude ($\simeq 2\pi$) in many cases, such as in
Fig.\ \ref{fig2}(b) and (c) at around 4T; in Fig.\ \ref{fig2}(c)
and (d) at around 2.5T, 4T and 8T; and at around 7.8T in \ref{fig2}(d) and
(e).

Stronger spin-tilting and occasional total flips appear close to the
region of nonzero (or $\neq 2\pi n$, with integer $n$) Berry phase,
as shown in Fig.\ \ref{fig2}(c) at around 2.5T, and in \ref{fig2}(d)
and (e) at around 7.8T. Thus, the spin hybridization and phase
modulation are intrinsically linked due to SOC and magnetic field.
Spin orientation and phase values smoothly change as a function of
magnetic field intensity (or magnetic field orientation in Fig.\
\ref{fig3}). Some apparently sudden spin-flips also appear, as the
one highlighted in panel \ref{fig2}(c), corresponding to a steep
(yet continuous) variation of the spin component, as detailed in the
inset. Similar smooth variations are presented for an eccentric
(elliptically deformed) ring on the right panels, Fig.\
\ref{fig2}(j)-(o). The main effect introduced by the confinement
asymmetry is to make the spin modulation and Berry phase vary more
gradually with field intensity. This can be understood as arising
from the asymmetry which introduces mixing of different angular
momentum components and associated anticrossings. Notice in Fig.
\ref{fig2}(i), that at higher magnetic fields, B$>$6T, various
levels mix. This can be seen in the large anticrossings between
levels 2 and 3 at around 7T, levels 4 and 5 at around 6.7T, and
levels 5 and 6 at around 8.5T.\@ The level mixture makes the
spectrum flatter with field and, correspondingly, produces weaker
variations in the Berry phase as well.

Figure \ref{fig2}(h) displays the gradual cumulative process
of adding Berry phases of the first 2, 4, and 6 consecutive levels of panels (b)-(g).
A similar addition has been obtained for
the asymmetric ring case, shown in Fig. \ref{fig2}(p). This additive process is equivalent to increasing
electron number or window around the Fermi level in a transport
experiment). The cumulative Berry phase, especially for large
number ($\gtrsim 3$) of levels counted is essentially null (or $=2\pi
n$). In fact, although individual levels show strong variation of
the Berry phase with field, the cumulative phase does not:
successive levels have compensating Berry phase changes, so that the
cumulative effect is surprisingly near null (except for occasional
$2\pi$ slips shown in the figure), especially for
the 4 and 6-level traces shown.

The introduction of quantum ring eccentricity changes the situation
in a somewhat subtle fashion. Comparing left (h) and right (p)
panels in Fig.\ \ref{fig2}, it is clear that as the eccentricity
induces changes in the electronic spectrum and single-state Berry
phases, the cumulative Berry phase shows gradual modulation, so that
nontrivial values are seen over finite-size windows in field: 2 - 3T and
7 - 9T for cumulative Berry phase of 2 levels; 0 - 1T and 6 - 7T,
for 4 levels; and 3 - 4T and 5 - 6T, for 6 levels. This would suggest that a moderate
degree of asymmetry and/or disorder unavoidably present in real
systems may in fact produce a more robust Berry phase signal in
experiments.

Similar contrasts exist between circularly symmetric and asymmetric
rings as function of magnetic field orientation
(at constant strength), as shown in Fig.\
\ref{fig3}. As in Fig.\ \ref{fig2}, each state shows a gradual Berry
phase evolution with magnetic field angle near level anticrossings, and the diamagnetic shift
provided by $B_z$ decreases for larger angles. One can also see a rather interesting
evolution of the spin orientation as the tilt angle increases. On the left panels, for the circularly
symmetric ring, one also notices relatively sharp changes in Berry
phase and spin orientation, as different angular momentum components
are mixed by spin orbit coupling.  Those jumps or drastic
changes disappear or become smoother for the asymmetric ring (right
panels), as the eccentricity mixes more strongly the different
angular momentum states. Panels (h) and (p) show the cumulative
Berry phase for the two rings. There is a similar behavior already
seen in Fig.\ \ref{fig2}: a smooth variation with angle for small
number of levels, changes to essentially null phase value ($2\pi n$)
for larger level number. The sudden phase slips
however become smoother, resulting in nontrivial values for the
asymmetric ring over wider range (angular in this case).

\begin{figure}[tbp]
\includegraphics[scale=1.0]{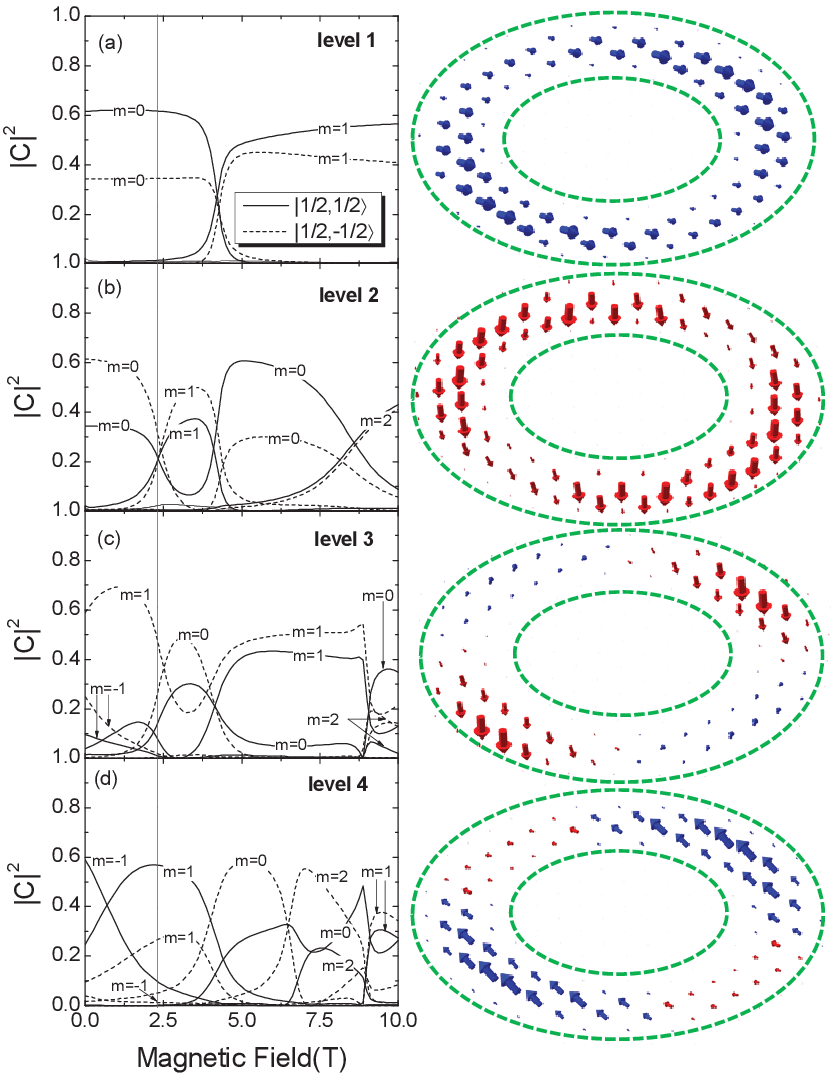}
\caption{Panels on the left column shown expansion coefficients for the four
lowest states of an asymmetric ring ($\delta=2$ meV), and fixed Rashba field $F=100$ kV/cm, and magnetic
field tilt angle $\theta=60^{\circ}$, as function of magnetic field intensity.  Level admixtures clearly
evolve with sudden switches at level anticrossings.
The right column shows spin density vector maps along the ring ($z$-integrated) for the
four lowest states at a field $B=2.375$ T (indicated by the vertical line in left panels).
Blue arrows have a positive projection along $z$, while
for red arrows the projection is negative.  Notice nearly parallel vectors in level 1 result in a null Berry phase;
in contrast, canting of vectors in level 2 contribute to a Berry phase of $\simeq -\pi$ (see
Fig.\ \protect\ref{fig2}(k)). } \label{fig4}
\end{figure}

The slow evolution of Berry phase for each state
signals the mixtures introduced by the different perturbations on
an otherwise highly-symmetric picture. The Zeeman field, SOC, and
structural asymmetries produce simultaneous mixtures of spin,
parity, and angular momentum. This effect, contained in the
expansion coefficients of the different states, can be visualized as
well through spin density maps.  Figure \ref{fig4}, left panels, show the
expansion coefficients for the four lowest energy states of an
asymmetric ring, as function of magnetic field, at a fixed angle
$\theta=60^{\circ}$. These panels show solid (dashed) curves for the
spin up (down) components with different angular momentum $m$ in the
given state.  The states are mixtures of angular momentum
(introduced at zero field by the ring asymmetry) and/or spin (due to SOC), which evolve with
field to other components (due to the diamagnetic shift of the
spectrum), and eventually to more complex mixtures at higher
energies.

The right panels in Fig.\ \ref{fig4} show spin vector maps for the
corresponding state at the field $B=2.375$T, and
$\theta=60^{\circ}$. This field value corresponds to the
anticrossing between the second and third levels in Fig.\
\ref{fig2}(i). The vector maps use arrows with size proportional to
the spin density at each point on the plane (integrating each
expression in Eq. (15) on the $z$-coordinate) and blue (or red)
colors to indicate a positive (or negative) sign of the $z$-spin
component at that point. The ground state (level 1) shows a spin map
predominantly on the plane, although with overall positive $S_z$
component, and with high amplitude near the ends of the long-axis
ellipse. The first-excited state, in contrast, shows large negative
$S_z$ components and with a spatial distribution that complements
that of the ground state. One also notices that the ground state
spin map shows local vectors that are essentially parallel all along
the ring: this would result in a vanishing Berry phase, as it is
indeed seen in Fig.\ \ref{fig2}(j), at this magnetic field. For the
second level, however, where the Berry phase $\simeq -\pi$ in Fig.\
\ref{fig2}(k), one notices that the spin arrows in Fig.\ \ref{fig4}
are canted with respect to those a quarter of the
way along the ring. It is this non-parallel nature of the spins
along the ring structure
that characterizes a non-vanishing Berry phase. Levels 3 and 4 show
even more structure, with spin vector amplitude more localized near
the long ends of the ellipse, but with $S_z$ component that changes
sign as one moves along the ring. The relative twisting of the spin
vector density along the ring, contributes to the non-vanishing
Berry phase seen in Fig. \ref{fig2}(l)-(m), although with much
smaller value than for level 2.  Other levels with non-vanishing
Berry phase show similar canted spin texture across the ring.

\section{Discussion and conclusions} \label{sec:disc}

We have used a nearly analytical description of the states in
quantum rings of finite width.  This model, used before to describe
realistic structures in experiments, allows us to extract
interesting insights on the role of spin-orbit coupling and its
interplay with external magnetic field effects, such as diamagnetic
shifts and Zeeman splitting.  We have moreover introduced asymmetry
in the confinement structure to see how this affects the level
structure and associated spin texture and Berry phase of different
states. We observed that possible experimental sweeps of magnetic
field tilt or amplitude, produce controllable changes in the state
characteristics, which can be traced in particular through the
smooth variation of the Berry phase of each state.  It is also clear
that as spin-orbit coupling could be made stronger with applied
electric fields, the Rashba effect would also
controllably change the overall geometric phase in
quantum rings.

Somewhat surprisingly, we found that the unavoidable defects or
asymmetries in ring confinement produce smooth changes in the Berry
phase, as either the magnetic field or tilt (or even Rashba field)
is changed.  This effect makes the otherwise sudden phase slips in
symmetric rings become smoother and produce non-vanishing (or
nontrivial) geometric phases as a consequence.  This would suggest
that moderate level mixing makes for more robust Berry phases in
experiments. One should also comment, that although the multilevel
cumulative Berry phase appears essentially null for higher level
number (or wider energy window), it may be possible to access
individual (or few) state Berry phases in narrow bias ranges or
similar other experiments where few states can be sampled.

\begin{acknowledgements}
The authors are grateful for financial support by CAPES-Brazil, CNPq-Brazil, and
MWN/CIAM NSF grant DMR-1108285. LKC is supported by FAPESP under grant 2012/13052-6.
\end{acknowledgements}

\end{document}